\documentclass[twocolumn,showpacs,preprintnumbers,amsmath,amssymb]{revtex4}

\usepackage{times}
\usepackage{epsfig}
\usepackage{graphicx}
\usepackage{dcolumn}
\usepackage{bm} 
\newcommand{\beq}{\begin{equation}}
\newcommand{\eeq}{\end{equation}}
\newcommand{\beqa}{\begin{eqnarray}}
\newcommand{\eeqa}{\end{eqnarray}}
\newcommand{\lam}{\lambda}

\newcommand{\ca}{{\bf\cal C}}
\newcommand{\ra}{\rangle}
\newcommand{\la}{\langle}

\def\ajp#1{{ Am.\ J.\ Phys.} {\bf #1}}

\def\jpb#1{{ J.\ Phys.\ B} {\bf#1}} 

\def\pr#1{{ Phys.\ Rev.} {\bf#1}} 
\def\pra#1{{ Phys.\ Rev. A\/} {\bf#1}}

\def\prl#1{{ Phys.\ Rev.\ Lett.} {\bf#1}}

\begin{document}

\title{Schmidt Analysis of Pure-State Entanglement}

\author{J.\ H.\ Eberly}
\email{eberly@pas.rochester.edu}
\affiliation{ Rochester Theory Center for Optical
Science and
Engineering\\
and \\
Department of Physics \& Astronomy, 
University of Rochester, Rochester, New York 14627 USA
}

\begin{abstract} We examine the application of Schmidt-mode analysis to pure state entanglement. Several examples permitting exact analytic calculation of Schmidt eigenvalues and eigenfunctions are included, as well as evaluation of the associated degree of entanglement.
\end{abstract}

\maketitle
\section*{Introduction} 
This paper is an overview of the Schmidt-mode approach \cite{Knight-Ekert} to the quantum description of entangled pure states, accompanied by several explicit examples in small-dimensional discrete spaces for illustration. These are presented in memory of Chuck Bowden, whose interest in direct, uncomplicated examinations of complex physical reality was one foundation of his skill as a teacher. 

Let us begin by considering the concrete example of a quantum state of two particles $|\Psi_{AB}\ra$. In the simplest situation to understand, this state is the product of states for each particle separately:
\beq \label{e.factored}
|\Psi_{AB}\ra = |\Psi_{A}\ra \otimes |\Psi_{B}\ra ,
\eeq
and then the state is obviously factored as it stands, in which case information about one of the particles is independent of information about the other. This means that extraction of information about particle $A$ by operations within the state space of particle $A$ will not change the content of state $|\Psi_{AB}\ra$ in the state space of $B$. Most multi-particle states are not of this type, and are such that extraction of information about particle $A$ does have an effect on the information available about other particles. 

This can be a subtle matter of correlation. Suppose that we know that the two particles are photons from a two-photon source tested to provide pairs of photons with opposite polarization in the horizontal-vertical basis, but without knowledge which photon (the green one or the red one, say) has which polarization. We can write the density matrix for these photons in two ways that equally express opposite HV polarizations of the photons:
\beqa 
\label{e.rhoCL}
\rho_{CL} &=& \frac{1}{2}\Big(|H_AV_B \ra \la H_AV_B|\Big) \nonumber \\
&+& \frac{1}{2}\Big(|V_AH_B\ra \la V_AH_B| \Big)  \\
\label{e.rhoQM}
\rho_{QM} &=& |\Psi(Bell)\ra \la \Psi(Bell)|, \quad {\rm where} \nonumber \\
|\Psi(Bell)\ra &=& \frac{1}{\sqrt2}\Big(| H_AV_B\ra + |V_AH_B\ra \Big) ,
\eeqa
where Bell state is a term reserved for the 4 basis states for two two-level quantum systems (i.e., of two qubits) \cite{BellState}.
For example, observation of H polarization for photon $A$, expressed by Trace$\{\rho |H_A\ra \la H_A|\}$ yields $|V_B\ra \la V_B|$, certain V polarization of $B$, in both cases. 

Regarding particles $A$ and $B$, the states represented by the two density matrices above are not entangled and entangled, respectively. The physical differences between them show up as two-particle coherences in the second case, coherences that are entirely quantum mechanical and not present in the first case, which has only classical correlations. 

An easy way to see that there will be such coherence differences is to display the $4\times4$ density matrices in the two cases:
\beq
\label{e.polnmatrices} 
\rho_{CL} = \frac{1}{2}\left(
\begin{array}{cccc}
0 & 0 & 0 & 0 \\
0 & 1 & 0 & 0 \\
0 & 0 & 1 & 0 \\
0 & 0 & 0 & 0
\end{array} \right)  ,\ 
\rho_{QM} = \frac{1}{2} \left(
\begin{array}{cccc}
0 & 0 & 0 & 0 \\
0 & 1 & 1 & 0 \\
0 & 1 & 1 & 0 \\
0 & 0 & 0 & 0
\end{array}  \right) ,\\
\eeq
where the rows and columns are ordered as HH, HV, VH, VV. Note that the non-zero off-diagonal elements, present only in $\rho_{QM}$, are in mixed-particle  positions, HV-VH and VH-HV, so they have no consequences for the particles individually, and each of these density matrices yields the same reduced density matrix for each particle separately, just 1/2 times the unit matrix. In experimental terms, the two photons in the second state will violate a Bell inequality, as was first demonstrated in 1972 in an experiment devised by Clauser \cite{Clauser}. The nature of a Clauser-type experiment that tests a Bell inequality has been described many times \cite{Eberly-AJP} and need not be repeated here.

\section*{A Specific Challenge}
\begin{figure}[!b]
\epsfig{file=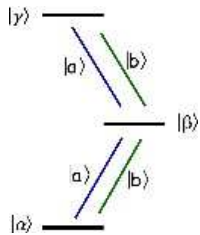, height=3cm}
\caption{\label{fig-energylev.eps} Atomic energy level diagram showing photon in one of two modes $|a\ra$ or $|b\ra$ interacting with three atomic states $|\alpha\ra, |\beta\ra$, and $|\gamma\ra$.}
\end{figure}

A different example gives an impression of the complexity of the issue when one tries to answer any question about ``degree" of entanglement. Let us consider an idealized scenario in which the two particles are an atom and a photon, as sketched in Fig. \ref{fig-energylev.eps}. Let the states of the atom be designated with Greek letters as $|\alpha\ra, |\beta\ra, |\gamma\ra, \dots$, and let the different photon modes be designated with Latin letters as $|m\ra, |n\ra, \dots$. Suppose that at the initial time $t=0$ the state $|\Psi_0\ra$ of the combined system is known to be given by:
\beqa \label{e.Psi0}
\sqrt{12}|\Psi_0\ra &=& \Big(2|a\ra + |b\ra \Big) \otimes |\alpha\ra \nonumber \\
&+&  \Big(|a\ra + 2|b\ra \Big) \otimes |\beta\ra \nonumber \\
&+&  \Big(|a\ra + |b\ra \Big) \otimes |\gamma\ra.
\eeqa
Is this an entangled state? It isn't easy at a glance to see how the state might be factored into a direct product of a complicated atom state and a complicated photonic state, but in the absence of contrary evidence such factoring is an open possibility. 

This illustrates that it is a common occurrence to be uncertain about entanglement, in this  case are the atom and photon entangled or not? They will turn out to be only slightly entangled, and one goal of this paper is to show how the Schmidt apporoach can answer both the entanglement question as well as interpret what ``slightly entangled" means for this case. More general questions, for example taking account of time evolution of entanglement, have interesting and challenging aspects that are still very incompletely understood \cite{Yu-Eberly}, but we will focus entirely on an issue that can be treated in detail, two-party entanglement at a fixed time  moment, and state (\ref{e.Psi0}) serves as a good example for attention. 

A different generalization to two-party entanglement in continuous spaces can be the basis for more additional discussion, including analyses of the Einstein-Podolsky-Rosen ``paradox" \cite{EPR}, but we will not pursue such matters. Recent work in this direction can be cited \cite{Law-Eberly04, Fedorov-etal04, Howell-Boyd05}.

\section*{Basis State Expansions in a Product Space}  

For a general two-particle pure state we can always write the state in terms of basis states for the two particles individually:
\beq \label{e.PsiAB}
|\Psi_{AB} \ra  = \sum_{i}\sum_{j} C(i,j) |i \ra \otimes |j \ra ,
\eeq
where the basis states are here denoted \{$|i\ra$\} and \{$|j\ra$\}. It is obvious that we can sensibly call the coefficient $C(i,j)$ the entanglement amplitude, since the two-particle state is clearly entangled unless the coefficient is factorable: $C(i,j) = P(i)Q(j)$. In our discussion it will helpful to have a notation that explicitly distinguishes the two spaces incorporated in the state, so we will follow the example in (\ref{e.Psi0}) above and rewrite the coefficient $C(i,j)$ as $C(n, \mu)$, where the Latin and Greek letters continue to serve the purpose of distinguishing the particle spaces. 

We proceed by treating $C(n, \mu)$ as the $n-\mu$ element of a matrix connecting the Latin and Greek states, which we will write ${\bf\cal C}$, so
\beq \label{e.CDef}
C(n, \mu) \equiv \la n| \ca |\mu \ra.
\eeq 
Note that $\ca$ converts a Greek state to a Latin one, and $\ca$ is generally not square, because the Latin and Greek dimensions need not be the same, as they aren't in $|\Psi_{0}\ra$ above. The adjoint matrix $\ca^\dag$ does the opposite. The products $\ca \ca^\dag$ and $\ca^\dag \ca$ operate entirely in the Latin and Greek spaces, respectively, and are easily seen to be Hermitean. It's an exercise to confirm that they have a direct physical interpretation as the reduced density matrices for the atom and photon: 
\beqa
\rho_L^{(0)} &\equiv& \sum_{\mu} \la \mu| \Big(|\Psi_{0}\ra \la \Psi_{0}| \Big) |\mu \ra = \ca\ca^\dag, \\
\rho_G^{(0)} &\equiv& \sum_n \la n| \Big(|\Psi_{0}\ra \la \Psi_{0} \Big) |n\ra = \ca^\dag\ca .
\eeqa

The reduced density matrices for the two particles play a key role because their eigenvalues and eigenvectors are important building blocks of any two-party quantum state. Let us write the eigenvalue equation for $\rho_L^{(0)}$ as
\beq \label{e.rhoL}
\rho_L^{(0)} |f^s \ra = \ca\ca^\dag|f^s\ra =  \lam_s |f^s\ra ,
\eeq  
where each $|f^s\ra$ eigenvector is defined to be normalized, and thus we have obtained a new orthonormal complete set of basis vectors in the Latin state space:
\beqa 
\label{e.orthonormf}
\la f^s|f^{r}\ra &=& \delta_{sr}  \quad {\rm and} \\
\label{e.completef}
\sum_s \la n| f^s\ra \la f^s|n'\ra &=& \delta_{nn'} .
\eeqa

Next we multiply (\ref{e.rhoL}) from the left by $\ca^\dag$, and add parentheses for clarity:
\beq
\ca^\dag\rho_L^{(0)} |f^s \ra = \ca^\dag\ca \Big(\ca^\dag|f^s\ra \Big) =  \lam_s \Big( \ca^\dag|f^s\ra \Big).
\eeq 
This shows that $\ca^\dag |f^s\ra$ is automatically an eigenvector of $\ca^\dag\ca = \rho_G^{(0)}$ in the Greek state space, with the same eigenvalue $\lam_s$:
\beq \label{e.rhoG}
\rho_G^{(0)} ~\Big(\ca^\dag|f^s\ra\Big) = \lam_s \Big(\ca^\dag|f^s\ra\Big). 
\eeq
We denote the new orthonormal set of Greek states by $|\phi^s\ra$, and they are defined by 
\beq \label{e.phiDef}
|\phi^s\ra \equiv \frac{1}{\sqrt\lam_s}\ca^\dag |f^s\ra,
\eeq
where the normalization factor $1/\sqrt\lam_s$ is required, as shown by the orthonormality condition $\la \phi^r| \phi^s\ra = \delta_{rs}$: 
\beqa
\la \phi^r| \phi^s\ra &=&  \la f^r|\ca \frac{1}{\sqrt{\lam_r}} \frac{1}{\sqrt{\lam_s}} \ca^\dag |f^s\ra = \frac{1}{\sqrt{\lam_s\lam_r}} \la f^r|\ca\ca^\dag|f^s\ra \nonumber\\
&=& \frac{1}{\sqrt{\lam_s\lam_r}} \lam_s\la f^r|f^s\ra = \delta_{rs}.
\eeqa

The adjoint relation to Eq. (\ref{e.phiDef}) is also useful for obtaining an explicit expression for the amplitude $C(n,\mu)$: $\ca|\phi^s\ra \equiv \frac{1}{\sqrt\lam_s}\ca\ca^\dag |f^s\ra 
= \sqrt{\lam_s}|f^s\ra$. From this we obtain
\beqa \label{e.componentsC}
\sqrt{\lam_s}\la n|f^s\ra &=& \la n|\ca |\phi^s\ra \nonumber \\
&=& \sum_{\nu}\la n|\ca |\nu\ra \la \nu|\phi^s\ra \nonumber \\
&=& \sum_{\nu}C(n, \nu) \la \nu|\phi^s\ra . 
\eeqa
Now we denote transformation coefficients between the original and the new basis states in both Latin and Greek spaces as:
\beq \label{e.coefficients}
f_n^s = \la n|f^s\ra, \quad {\rm and} \quad \phi_\nu^s = \la \nu|\phi^s \ra
\eeq
and then the first and last lines above, when multiplied by $\la\phi^s|\mu\ra$ and summed over $s$, allow the orthonormality of the $|\phi^s\ra$ states to give
\beq \label{e.CSoln}
C(n, \nu) = \sum_s \sqrt\lam_s f_n^s(\phi_\nu^s )^*,
\eeq
which is the desired expression for the entanglement amplitude.

\section*{Schmidt Modes for Two-Particle Pure States}  

The derivation just given allows the original state to be written in the so-called Schmidt form, as follows:
\beqa \label{e.SchmidtPsi}
|\Psi_{0}\ra &\equiv& \sum_n \sum_\nu C(n, \nu) |n\ra \otimes |\nu\ra \nonumber \\
&=& \sum_n \sum_\nu \Big(\sum_s \sqrt\lam_s f_n^s (\phi^s_{\nu})^* \Big) |n\ra \otimes |\nu\ra \nonumber \\
&=& \sum_s \sqrt\lam_s \Big(\sum_n f^s_n |n\ra \Big) \otimes \Big( \sum_\nu (\phi^s_{\nu})^* |\nu\ra \Big) \nonumber \\
&=& \sum_s \sqrt\lam_s |F^s\ra \otimes |\Phi^s\ra , 
\eeqa
where the vectors $|F^s\ra$ and $|\Phi^s\ra$ can be called the Schmidt modes for the state, with expressions in terms of the $|n\ra$ and $|\nu\ra$ basis states as defined in the last line. It is worth a specific remark that the original mode expression for $|\Psi_{0}\ra$ given in formula (\ref{e.Psi0}) required a double summation, whereas the Schmidt form requires only a single summation. 

It is clear from the concept of state factorization, and the fact that the Schmidt states are orthogonal bases, that a state cannot be factored, i.e. is entangled, if more than a single term is needed in its Schmidt sum. Also, since the $\lam$s are eigenvalues of density matrices, they can reasonably be termed ``information eigenvalues." Naturally, since we must have $1 \ge \lam_s \ge 0$ as well as  $\sum_s \lam_s = 1$, if one of the eigenvalues equals 1, the rest must be zero, meaning that there is only a single term in the Schmidt sum and the state is factored as it stands in (\ref{e.SchmidtPsi}). 

\begin{figure}[!b]
\epsfig{file=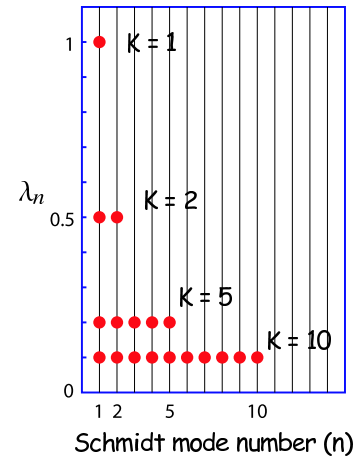, height=6cm}
\caption{\label{fig-Kchart.eps} Reduced density matrix eigenvalues in the special case that all the eigenvalues are equal, with associated $K$ values for situations of different dimensionality.}
\end{figure}

Following in the same line, an important possibility, given the Schmidt form, is to define a ``degree" of entanglement or a quantitative measure of ``how much entangled" the state is.  This could be done by counting the number of terms in the Schmidt sum, since having more terms means that information is more ``spread out" and so multiple projective measurements of one particle are required to get definite information about the other. However, if many of the eigenvalues $\lambda_s$ are very small, their terms contribute little to the sum, so a weighted measure is preferable to a simple count of eigenvalues, as is sometimes advocated \cite{Nielsen-Chuang}. Logarithmic measures related to entropy can be cited for a weighted degree of entanglement, but the simplest measure for pure states takes squared eigenvalues as the weights, and one set of examples is plotted in Fig. \ref{fig-Kchart.eps}.

The so-called Schmidt number $K$, as we use it, is defined as \cite{Grobe-etal95}:
\beq \label{e.Schmidtno}
K \equiv \frac{1}{Tr_A \{(\rho^A)^2\}} \equiv \frac{1}{Tr_B \{(\rho^B)^2\}}\equiv \frac{1}{\sum \lambda_n^2},
\eeq
which is the reciprocal of the purity of either one of the reduced states. This is sensible to interpret as a count of the ``effective" number of terms in the sum (\ref{e.SchmidtPsi}) as follows. First, if one $\lam$ is 1 and the rest zero, the definition gives $K=1$. Then, at the other extreme, suppose that all $\lambda$s are equal, meaning that all $N$ terms are equally important in the Schmidt expression (\ref{e.SchmidtPsi}) for the state. Clearly, if there are $N$ $\lam$s, one then has $\lam_s = 1/N$ for all $s$, and the consequence is $K = N$, accurately counting all the equal-weighted terms. In any other arrangement of $\lam$s the number is smaller, so if $D$ is the dimension of the state space, then  $D \ge K \ge 1$. As an exercise one can check that for all of the Bell polarization states \cite{BellState} we have $K = 2$, the maximum possible value when only two-dimensional polarization states are involved. The $K$ measure has a potentially direct experimental relevance as it tells how many distinct modes the experimenter must expect to be needed.

\section*{Original Example in Schmidt Form}

Now we can return to formula (\ref{e.Psi0}) for a two-particle (atom-photon) state. It is a short exercise to determine the matrix form of the entanglement amplitude:
\beq \label{e.Cmatrices}
\ca = \sqrt{\frac{1}{12}} \left(
\begin{array}{ccc}
2 & 1 & 1\\
1 & 2 & 1 
\end{array} \right)~, \quad 
\ca^\dag = \sqrt{\frac{1}{12}}\left(
\begin{array}{cc}
2 & 1 \\
1 & 2 \\
1 & 1 
\end{array} \right) ,
\eeq
so that the reduced Greek and Latin density matrices are given  by:
\beqa
\label{e.reducedmatrices} 
\rho_G^{(0)} &=& \ca^\dag\ca = \frac{1}{12}\left(
\begin{array}{ccc}
5 & 4 & 3 \\
4 & 5 & 3 \\
3 & 3 & 2
\end{array} \right) \\
&\quad& \ \nonumber \\
\rho_L^{(0)} &=& \ca\ca^\dag = \frac{1}{12} \left(
\begin{array}{cc}
6 & 5\\
5 & 6
\end{array}  \right) \nonumber \\ 
&\to& \frac{1}{12} \left(
\begin{array}{ccc}
6 & 5 & 0 \\
5 & 6 & 0 \\
0 & 0 & 0
\end{array}  \right),\\
\eeqa
where the final expression is only a formal change to match the dimensions of the density matrices, for easier manipulation. It can even be given a physical interpretation since it is the same as including a third mode of the radiation field, one that has no interaction with the atom.

\begin{figure}[!b]
\epsfig{file=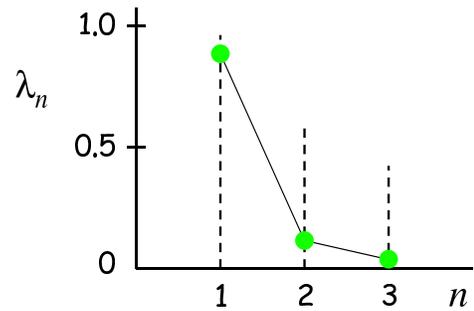, height=4cm}
\caption{\label{fig-eigen.eps} Eigenvalues plotted for either $\rho_L^{(0)}$ or $\rho_G^{(0)}$. The fact that $\lam_1 \approx 1$  means that only the first term in the Schmidt sum is very important, so the state cannot be highly entangled.}
\end{figure}

The additional row and column in $\rho_L^{(0)}$ guarantee that it will have zero as an eigenvalue. We have claimed that $\rho_L^{(0)}$  and $\rho_G^{(0)}$  have the same eigenvalues, and it is not hard to show that in this case the eigenvalues are:
\beq \label{e.eigenvalues}
\lam = \frac{11}{12},\ \frac{1}{12},\ 0
\eeq
for both matrices, as plotted in Fig. \ref{fig-eigen.eps}. From the $\lam$s  we easily compute the Schmidt number as the degree of entanglement:
\beq \label{Kvalue}
K = \frac{1}{\sum_s \lam_s^2} = \frac{(12)^2}{(11)^2 + 1^2 + 0}
= \frac{144}{122} \approx 1.18,
\eeq
which is close enough to 1 to justify the earlier remark that $|\Psi_0\ra$ is very little entangled.

The eigenvectors for the two density matrices are not the same, and those for the non-zero Latin eigenvalues can be shown to be
\beqa \label{e.Lvectors}
|F^{(11)}\ra &=&  \frac{1}{\sqrt2}\left( |a\ra + |b\ra \right),\\
|F^{(1)}\ra &=&  \frac{1}{\sqrt2}\left( |a\ra - |b\ra \right),
\eeqa 
and those for the Greek states with non-zero eigenvalues follow by application of relation (\ref{e.phiDef}):
\beqa \label{e.Gvectors}
|\Phi^{(11)}\ra &=& \frac{1}{\sqrt{22}} \left(3|\alpha\ra + 3|\beta\ra + 2|\gamma\ra \right), \\
|\Phi^{(1)}\ra &=& \frac{1}{\sqrt2}\left( |\alpha\ra - |\beta\ra \right),
\eeqa 
where the orthnormality of these states should be checked. Now the original state can be written in the Schmidt basis as:
\beqa \label{e.PsiSchmidt}
\sqrt{12}~|\Psi_0\ra &=& \sqrt{{11}}~|\Phi^{(11)}\ra \otimes |F^{(11)}\ra \nonumber \\
&+& \sqrt1~|\Phi^{(1)}\ra \otimes |F^{(1)}\ra \nonumber \\
&=& \frac{1}{2}\Big(3|\alpha\ra + 3|\beta\ra + 2|\gamma\ra \Big) \otimes \Big( |a\ra + |b\ra \Big) \nonumber \\ 
&+&  \frac{1}{2}\Big( |\alpha\ra - |\beta\ra \Big) \otimes \Big( |a\ra - |b\ra \Big) , 
\eeqa
which can easily be rearranged into the form originally given in (\ref{e.Psi0}).

\section*{Some More Examples}

The same procedure applies to all two-particle pure states, without regard to dimension. Several examples serve as additional illustrations. Very small changes can have relatively large consequences. For instance, if we reproduce the original $|\Psi_0\ra$ in (\ref{e.Psi0}), but change a single sign in the last term, we get:
\beqa \label{e.Psi1}
\sqrt{12}|\Psi_1\ra &=& \Big(2|a\ra + |b\ra \Big) \otimes |\alpha\ra \nonumber \\
&+& \Big(|a\ra + 2|b\ra \Big) \otimes |\beta\ra \nonumber \\
&+& \Big(|a\ra - |b\ra \Big) \otimes |\gamma\ra.
\eeqa
In this case the transformation matrix is given by
\beq
\label{e.C1} 
\ca^{(1)} =  \sqrt{\frac{1}{12}}\left(
\begin{array}{ccc}
2 & 1 & 1 \\
1 & 2 & -1 \\
\end{array} \right) .
\eeq
and the reduced density matrix for the Greek state space reads:
\beq
\label{e.reducedmatrices1} 
\rho_G^{(1)} = \ca^{(1)\dag}\ca^{(1)} = \frac{1}{12}\left(
\begin{array}{ccc}
5 & 4 & 1 \\
4 & 5 & -1 \\
1 & -1 & 2
\end{array} \right) .
\eeq
Its eigenvalues are 
\beq
\lam^{(1)} = \frac{3}{4}, ~\frac{1}{4}, ~0,
\eeq
and these indicate a Schmidt number $K = 1.6$, corresponding to a significant amount of entanglement for a two-eigenvalue problem. It is an easy exercise to find the Schmidt eigenvectors.

Higher dimensional spaces are treated in the same way.  An alternative photon-atom state to consider is one that involves an additional atomic state $|\delta\ra$:
\beqa \label{e.Psi2}
N_2|\Psi_2\ra &=& \Big(2|a\ra + |b\ra \Big) \otimes |\alpha\ra + \Big(|a\ra + 2|b\ra \Big) \otimes |\beta\ra  \nonumber \\
&+& \Big(|a\ra - |b\ra \Big)\otimes \Big(|\gamma\ra  - |\delta\ra  \Big),
\eeqa
where the overall normalization needs to be determined. The Greek density matrix is here $4\times4$:
\beq
\rho_G^{(2)} = \frac{1}{N_2^{2}}\left(
\begin{array}{cccc}
5 & 4 & 1 & -1 \\
4 & 5 & -1 & 1 \\
1 & -1 & 2 & -2 \\
-1 & 1 & -2 & 2
\end{array} \right),
\eeq
and the Schmidt number can be found to be higher than in either previous example. An easy question for the reader: how many non-zero eigenvalues are there in this case (before making any calculations of them)? Another exercise is to determine the transformation matrices $\ca$ and $\ca^\dag$ for this example.

\section*{Summary}

We have given a didactic review of the little-used Schmidt-mode approach to two-particle pure states, with the goal of increasing its familiarity. The transformation matrix $\ca$ was shown to play a central role. We explained how application of the Schmidt decomposition automatically provides a straightforward test of entanglement, and includes calculation of ``information eigenvalues" that lead to an easily-applied definition of degree of entanglement denoted $K$. The review has been illustrated by several explicit examples and contains several questions left as challenges. A slightly more subtle challenge for the interested reader is posed when some of the coefficients are complex, as in the example:
\beqa \label{e.Psi3}
N_3|\Psi_3\ra &=& \Big(2|a\ra + i|b\ra \Big) \otimes |\alpha\ra \nonumber \\
&+&  \Big(i|a\ra + 2|b\ra \Big) \otimes |\beta\ra \nonumber \\
&+&  \Big(|a\ra + |b\ra \Big) \otimes |\gamma\ra.
\eeqa
In this case one can find that the state is almost as entangled as possible, with $K = \frac{144}{74} \approx 2$.

\section*{Acknowledgements}

This review reports results related to research initiated under NSF Grant PHY 00-72359. It is intended as a small tribute to Chuck Bowden and his passion for clear science and science that has a purpose. He learned of the details of the Schmidt decomposition and its utility in pure state entanglement study during a NATO Institute in Turkey and was fascinated by the possibility of its applications. I thank Drs. H. Huang and K.W. Chan and Prof. C.K. Law for a number of useful remarks concerning Schmidt analysis.


\begin{thebibliography}{99}
\bibitem{BellState} Bell State is the term given to any of the four states $|\Psi^{\pm}\ra = \frac{1}{\sqrt2}\Big(| H_A\ra \otimes |V_B\ra  \pm |V_A\ra \otimes |H_B \ra \Big)$ and $|\Phi^{\pm}\ra \equiv \frac{1}{\sqrt2}\Big(| H_A\ra \otimes |H_B\ra  \pm |V_A\ra \otimes |V_B \ra \Big)$.
\end{thebibliography}
\end{document}